\documentclass [12pt,a4paper]{article}
\usepackage{amssymb, theorem}
\usepackage{graphicx}
\usepackage{graphics}
\newtheorem{Thm}{Theorem}
\newtheorem{lemma}[Thm]{Lemma}
\theorembodyfont{\rm}

\def\proof{{\it Proof: }}

\def\qed{\nobreak\hfill $\square$}
\def\<{\langle}
\def\>{\rangle}
\def\bal{\langle\!\langle}
\def\jobb{\rangle\!\rangle}

\def\J{{\mathbb J}}
\def\bC{{\mathbb K}}
\def\bJ{{\mathbb J}}
\def\bL{{\mathbb L}}
\def\bR{{\bf R}}

\def\iK{{\cal K}}

\def\bL{{\bf L}}

\def\im{\mathrm{i}}

\def\bbbr{{\mathbb R}}

\def\Diag{\mbox{Diag}\,}
\def\Cov{\mathrm{Cov}}
\def\QCov{\mathrm{qCov}}

\def\Det{\mathrm{Det}\,}

\def\Tr{\mathrm{Tr}\,}
\def\pont{{\, \cdot \,}}

\begin{document}
 \ \vskip 1cm 
\centerline{\large {\bf Quantum covariance, quantum Fisher information}}
\bigskip
\centerline{\large {\bf and the uncertainty principle}}
\bigskip
\bigskip\bigskip
\centerline{\large 
Paolo Gibilisco\footnote{E-mail:  
gibilisco@volterra.uniroma2.it}$^{,4}$,
Fumio Hiai\footnote{E-mail: hiai@math.is.tohoku.ac.jp. 
Partially supported by Grant-in-Aid for Scientific 
Research (B)17340043.}$^{,5}$ and
D\'enes Petz\footnote{E-mail: petz@math.bme.hu.
Partially supported by the Hungarian Research Grant OTKA T068258
and 49835.}$^{,6}$
}
\bigskip
\begin{center}
$^4$ Dipartimento SEFEMEQ, Universit\`a di Roma ``Tor Vergata", \\ 
Via Columbia 2, 00133-Rome, Italy.
\end{center}

\begin{center}
$^5$ Graduate School of Information Sciences, Tohoku University \\
Aoba-ku, Sendai 980-8579, Japan
\end{center}

\begin{center}
$^6$ Alfr\'ed R\'enyi Institute of Mathematics, \\H-1364 Budapest,
POB 127, Hungary
\end{center}

\begin{abstract}
In this paper the relation between quantum covariances and quantum Fisher 
informations are studied. This study is applied to generalize a recently proved 
uncertainty relation based on quantum Fisher information. The proof given here 
considerably simplifies the previously proposed proofs and leads to more general
inequalities.
\smallskip


\end{abstract}

\bigskip
\bigskip\bigskip

\section{Introduction}
Fisher information has been an important concept in mathematical statistics
and it is an ingredient of the Cram\'er-Rao inequality. It was extended
to a quantum mechanical formalism in the 1960's by Helstrom \cite{He}
and later by Yuen and Lax \cite{YL}, see \cite{Ho} for the rigorous version.

The state of a finite quantum system is described by a density matrix 
$D$ which is positive semi-definite with $\Tr D =1$. If $D$ depends on a 
real parameter $-t <\theta < t$, then the true value of $\theta$ can be 
estimated by a self-adjoint matrix $A$, called observable, such that
$$
\Tr D_\theta A= \theta.
$$
This means that expectation value of the measurement of $A$ is the true
value of the parameter (unbiased measurement). When the measurement is 
performed (several times on different copies of the quantum system), the 
average outcome is a good estimate for the parameter $\theta$.

It is convenient to choose the value $\theta=0$. Then the  Cram\'er-Rao 
inequality has the form
$$
\Tr D_0 A^2 \ge \frac{1}{\mbox{Fisher information}},
$$
where the Fisher information quantity is determined by the parametrized
family $D_\theta$ and it does not depend on the observable $A$, see 
\cite{Ho, qinfobook}

The Fisher information depends on the tangent of the curve $D_\theta$.
There are many curves through the fixed $D_0$ and the Fisher information
is defined on the tangent space. The latter is the space of traceless 
self-adjoint matrices in case of the affine parametrization of the state 
space. The Fisher information is a quadratic form depending on the foot
point $D_0$. If it should generate a Riemannian metric, then it should
depend on $D_0$ smoothly \cite{Amari}.

\section{From coarse-graining to Fisher information and covariance}

Heuristically, coarse-graining implies loss of information, therefore
Fisher information should be monotone under coarse-graining. This was proved
in \cite{Ce} in probability theory and a similar approach was proposed in
\cite{Mo-Ce} for the quantum case. The approach was completed in \cite{PD2},
where a class of quantum Fisher information quantities was introduced, see
also \cite{PD22}.

Assume that $D_\theta$ is a smooth curve of density matrices with
tangent $A:=\dot{D}_0$ at $D_0$. The quantum Fisher information 
$F_D(A)$ is an information quantity associated with the pair $(D_0, A)$
and it appeared in the Cram\'er-Rao inequality above. Let now $\alpha$ be 
a coarse-graining, that is $\alpha : M_n \to M_k$ is a completely positive 
trace-preserving mapping. Then $\alpha (D_\theta)$ is another curve in $M_k$. 
Due to the linearity of  $\alpha$, the tangent at $\alpha(D_0)$ is 
$\alpha(A)$. As it is usual in statistics, information cannot be gained 
by coarse graining, therefore we expect that the Fisher information at 
the density matrix $D_0$ in  the direction $A$ must be larger than the 
Fisher information at $\alpha(D_0)$ in the direction $\alpha(A)$. 
This is the monotonicity property of the Fisher information under 
coarse-graining:
\begin{equation}\label{E:FM}
F_D(A) \ge F_{\alpha(D)}(\alpha(A))
\end{equation}

Another requirement is that $F_D(A)$ should be quadratic in $A$, in
other words there exists a (non-degenerate) real positive bilinear form 
$\gamma_D(A,B)$ on the self-adjoint matrices such that
\begin{equation}\label{E:BFM}
F_D(A)=\gamma_D(A,A).
\end{equation}
The requirements (\ref{E:FM}) and (\ref{E:BFM}) are strong enough
to obtain a reasonable but still wide class of possible quantum Fisher
informations. 

The bilinear form $\gamma_D(A,B)$ can be canonically extended to the
positive sesqui-linear form (denoted by the same $\gamma_D$) on the 
complex matrices, and we may assume that 
$$
\gamma_D(A,B)=\Tr A^*\J_D^{-1}(B)
$$
for an operator $\J_D$ acting on matrices. (This formula
expresses the inner product $\gamma_D$ by means of the
Hilbert-Schmidt inner product and the positive linear 
operator $\J_D$.) Note that this notation transforms (\ref{E:FM})
into the relation
$$
\alpha^* \bJ_{\alpha(D)}^{-1} \alpha \le \bJ_{D}^{-1},
$$
which is equivalent to
\begin{equation}\label{E:fimon1}
\alpha \J_{D}\alpha^*  \le \J_{\alpha(D)}\, .
\end{equation}

Under the above assumptions, there exists a unique operator
monotone function $f: \bbbr^+\to \bbbr$ such that $f(t)=tf(t^{-1})$
and 
\begin{equation}\label{E:Jdef3}
\J_D=f(\bL_D\bR_D^{-1})\bR_D\,,
\end{equation}
where the linear transformations $\bL_D$ and  $\bR_D$ acting on matrices
are the left and right multiplications, that is
$$
\bL_D(X)=DX \qquad\mbox{and}\qquad \bR_D(X)=XD\,.
$$
To be adjusted to the classical case, we always assume that $f(1)=1$
\cite{PD2, PD3}. It seems to be convenient to call a function $f:\bbbr^+ \to 
\bbbr^+ $ {\bf standard} if $f$ is operator monotone, $f(1)=1$ and
$f(t)=tf(t^{-1})$. (A standard function is essential in the context of
operator means \cite{Ando, PD2}.)

If $D=\Diag(\lambda_1, \lambda_2,\dots, \lambda_n)$ (with $\lambda_i >0$), 
then
\begin{equation}\label{E:Jdef2}
\gamma_D(A,B)=\sum_{ij} \frac{1}{M_f(\lambda_i,\lambda_j)}\overline{A}_{ij}
B_{ij},
\end{equation}
where $M_f$ is the mean induced by the function $f$:
$$
M_f(a,b):=bf(a/b).
$$
When $A$ and $B$ are self-adjoint, the right-hand-side of (\ref{E:Jdef2})
is real as required since $M_f(a,b)=M_f(b,a)$.

Similarly to Fisher information, the covariance is a bilinear form as well.
In probability theory, it is well-understood but the non-commutative 
extension is not obvious. The monotonicity under coarse-graining should hold:
\begin{equation}\label{E:Cov}
\QCov_D(\alpha^*(A), \alpha^*(A)) \le \QCov_{\alpha(D)}(A, A),
\end{equation}
where $\alpha^*$ is the adjoint with respect to the Hilbert-Schmidt inner
product. ($\alpha^*$ is a unital completely positive mapping.) If 
the covariance is expressed by the Hilbert-Schmidt inner
product as
$$
\QCov_{D}(A, B)=\Tr A^* \bC_D(B),
$$
then the monotonicity (\ref{E:Cov}) has the form
$$
\alpha \bC_D \alpha^* \le \bC_{\alpha(D)}.
$$
This is actually the same relation as (\ref{E:fimon1}). Therefore,
condition (\ref{E:Cov}) implies 
$$
\QCov_D(A,B)=\Tr A^*\J_D(B),
$$
where $\J_D$ is defined by (\ref{E:Jdef3}). The one-to-one correspondence 
between Fisher information quantities and (generalized) covariances was
discussed in \cite{PD22}. The analogue of formula (\ref{E:Jdef2}) is
\begin{equation}\label{E:Jdef5}
\QCov_D(A,B)=\sum_{ij} M_f(\lambda_i,\lambda_j)\overline{A}_{ij}B_{ij}
-\Big(\sum_{i} \lambda_i \overline{A}_{ii}\Big)
\Big(\sum_{i} \lambda_i B_{ii}\Big).
\end{equation}

If we want to emphasize the dependence of the Fisher information and the 
covariance on the function $f$, we write $\gamma^f_D$ and $\QCov^f_D$.
The usual symmetrized covariance corresponds to the function $f(t)=(t+1)/2$:
$$
\QCov^f_{D}(A,B)=\Cov_D(A,B):=
\frac{1}{2}\Tr (D(A^*B+BA^*))- (\Tr DA^*)(\Tr DB)
$$
Of course, if $D, A$ and $B$ commute, then $\QCov^f_D(A,B)=\Cov_{D}(A,B)$
for any standard function $f$. Note that both $\QCov^f_D$ and 
$\gamma^f_D$ are particular quasi-entropies \cite{OP, PD32}.

\section{Relation to the commutator}

Let $D$ be a density matrix and $A$ be self-adjoint. The commutator $\im 
[D,A]$ appears in the discussion about Fisher information. One
reason is that the tangent space $T_D:=\{B=B^*: \Tr DB=0\}$ has 
a natural orthogonal decomposition:
$$
\{B=B^*: [D,B]=0\}\oplus \{\im[D,A]:A=A^*\}.
$$

For self-adjoint operators $A_1, ...,A_N$, Robertson's uncertainty
principle is the inequality
$$
\Det \Big[\Cov_{D}(A_i,A_j)\Big]_{i,j=1}^N 
\ge
\Det \Big[ - \frac{\im}{2} \Tr D [A_i,A_j]\Big]_{i,j=1}^N ,
$$
see \cite{Rob}. The left-hand side is known in classical probability as 
the generalized variance of the random vector $(A_1, ..., A_N)$.
A different kind of uncertainty principle has been recently conjectured in \cite{GII} 
and proved in \cite{GII:2007a, andai}:  
\begin{equation}\label{dyn}
\Det \Big[\Cov_D (A_i,A_j)\Big]_{i,j=1}^N  
\ge
\Det \Big[ \frac{f(0)}{2} \gamma_D^f(
\im[D,A_i], \im[D, A_j])\Big]_{i,j=1}^N.
\end{equation}
Particular cases of inequality (8) have been proved in \cite{GI,3, H, LZ1, LZ2, LZ3, 
Kosaki, YFK}.
Of course, we have a non-trivial inequality in the case $f(0)>0$. The 
inequality can be called {\bf dynamical uncertainty principle}, since
the right-hand-side is the volume of a parallelepiped  determined
by the tangent vectors of the trajectories of the time-dependent 
observables $A_i(t):=D^{\im t}A_i D^{-\im t}$. Another remarkable
property is that inequality (\ref{dyn}) gives a non-trivial bound also 
in the odd case $N=2m+1$ and this seems to be the first result of this 
type in the literature.

The right-hand-side of (\ref{dyn}) is Fisher information
of commutators. If
\begin{equation}\label{E:tilde1}
\tilde{f}(x):=\frac{1}{2}\left((x+1)-(x-1)^2 \frac{f(0)}{f(x)}\right),
\end{equation}
then
\begin{equation}\label{E:tilde2}
\frac{f(0)}{2} \gamma_D^f(\im [D,A], \im [D,B] )=
\Cov_D(A,B)-\QCov^{\tilde f}_D (A, B )
\end{equation}
for $A,B \in T_D$. Identity (\ref{E:tilde2}) is easy to check but it is
not obvious that for a standard $f$ the function $\tilde{f}$ is operator 
monotone. It is indeed true that $\tilde{f}$ is a standard function as well, 
see Propositions 5.2 and 6.3 in \cite{3}. Note that the left-hand-side of 
(\ref{E:tilde2}) was called (metric adjusted) skew information in \cite{H}. 

\section{Inequalities}

In this section we give a simple new proof for the dynamical uncertainty
principle (\ref{dyn}). The new proof actually gives a slightly more 
general inequality.
 
\begin{Thm}\label{T:1}
Assume that $f,g:\bbbr^+\to \bbbr$ are standard functions such that
\begin{equation}\label{E:felt1}
g(x)\ge c \frac{(x-1)^2}{f(x)}
\end{equation}
for some $c>0$. Then
$$
\QCov_D^g (A,A) \ge c\, \gamma^f_D ([D,A], [D,A]).
$$
\end{Thm}

\proof
We may assume that $D=\Diag(\lambda_1, \lambda_2,\dots, \lambda_n)$
and $\Tr DA=0$. Then the left-hand-side is
$$
\sum_{ij} M_g(\lambda_i,\lambda_j) |A_{ij}|^2
$$
while the right-hand-side is
$$
c\sum_{ij} \frac{(\lambda_i-\lambda_j)^2}{M_f(\lambda_i,\lambda_j)} |A_{ij}|^2.
$$
The proof is complete. \qed

For any standard function $f$ and its transform $\tilde{f}$ given by
(\ref{E:tilde1}), $\tilde{f}\ge 0$ is exactly
$$
\frac{1+x}{2}-\frac{f(0)(x-1)^2}{2f(x)} \ge 0.
$$
Therefore for $g(x)=(1+x)/2$ the assumption (\ref{E:felt1}) holds for any $f$
if $c=f(0)/2$. Actually, this is the point where the operator monotonicity
of $f$ is used, in Theorem \ref{T:1} only inequality (\ref{E:felt1})
was essential.

The next lemma is standard but the proof is given for completeness.

\begin{lemma}\label{L:2}
Let $\iK$ be a finite dimensional real Hilbert space with inner product
$\bal \pont , \pont \jobb$. Let $\< \pont ,\pont\>$ be a real
{\rm(}not necessarily strictly{\rm)} positive bilinear form on $\iK$. If
$$
\< f,f\> \le \bal f, f \jobb
$$
for every vector $f \in \iK$, then 
\begin{equation}\label{E:det}
\Det \left(\,[\, \< f_i,f_j\>\,]_{i,j=1}^m\,\right) \le 
\Det \left(\, [\,\bal f_i, f_j \jobb \,]_{i,j=1}^m\,\right)
\end{equation}
holds for every $f_1,f_2,\dots, f_m \in \iK$. Moreover, if
$\bal \pont , \pont \jobb - \< \pont ,\pont\>$ is strictly positive, then inequality
{\rm(\ref{E:det})} is strict whenever $f_1,\dots,f_m$ are linearly independent.
\end{lemma}

\proof
Consider the Gram matrices $G:=[\,\bal f_i, f_j \jobb \,]_{i,j=1}^m$ and
$H:=[\, \< f_i,f_j\>\,]_{i,j=1}^m$, which are symmetric and
positive semidefinite. For every $a_1,\dots,a_m\in\bbbr$ we get
$$
\sum_{i,j=1}^m (\bal f_i, f_j \jobb - \< f_i,f_i\>) a_ia_j
= \bal \sum_{i=1}^m a_if_i, \sum_{i=1}^m a_if_i \jobb
- \< \sum_{i=1}^m a_if_i, \sum_{i=1}^m a_if_i \> \ge 0
$$
by assumption. This says that $G-H$ is positive semidefinite, hence it is clear that
$\Det(G)\ge\Det(H)$.

Moreover, assume that $\bal \pont , \pont \jobb - \< \pont ,\pont\>$ is strictly
positive and $f_1,\dots,f_m$ are linearly independent. Then $G-H$ is positive definite
and hence $\Det(G)>\Det(H)$.\qed

The previous general result is used now to have a determinant inequality,
an extension of the dynamical uncertainty relation.

\begin{Thm}\label{T:3}
Assume that $f,g:\bbbr^+\to \bbbr$ are standard functions such that
$$
g(x)\ge c \frac{(x-1)^2}{f(x)}
$$
for some $c>0$. Then for self-adjoint matrices 
$A_1,A_2, \dots ,A_m$ the determinant inequality
\begin{equation}\label{E:fo}
\Det\biggl( \left[\QCov_{D}^g(A_i,A_j)\right]_{i,j=1}^m \biggr) \ge
\Det\biggl( \left[ c\,\gamma_D^f\left( \,[D,A_{i}], [D,A_{j}]\, \right)
\right]_{i,j=1}^m  \biggr)
\end{equation}
holds.  
Moreover, equality holds in {\rm(\ref{E:fo})}  if and only if
$A_i-(\Tr DA_i)I$, $1\le i\le m$, are linearly dependent, and both sides of
{\rm(\ref{E:fo})} are zero in this case.
\end{Thm}

\proof
Let $\iK$ be the real vector space $T_D=\{ B=B^*: \Tr D B =0\}$.
We have $\QCov_{D}^g(A,A)=0$ if and only if $A=\lambda I$, therefore 
$$
\bal A, B \jobb :=\QCov_{D}^g(A,B)
$$
is an inner product on $\iK$. From formulas (\ref{E:Jdef2}),
(\ref{E:Jdef5}) and from the hypothesis, we have
\begin{eqnarray*}
c\gamma_D^f\left( \,[D,A], [D,A]\, \right)& = &
\sum_{ij} c \frac{(\lambda_i -\lambda_j)^2}{M_f(\lambda_i , \lambda_j)}
|A_{ij}|^2 \cr & \le & \sum_{ij} M_g(\lambda_i , \lambda_j)|A_{ij}|=
\QCov_{D}^g(A,A)=\bal A, A \jobb.
\end{eqnarray*}
If 
$$
\< A,B\>:= c\gamma_D^f\left( \,[D,A], [D,B]\, \right),
$$
then $\< A,A\> \le \bal A, A \jobb$ holds and (\ref{E:det}) gives the
statement when $\Tr DA_1=\Tr DA_2=\dots=\Tr DA_m=0$. The general case
follows by writing $A_i -(\Tr DA_i) I$ in place of $A_i$, $1 \le i \le m$.

To prove the statement on equality case, we show that $g(x) > c(x-1)^2/f(x)$ or
$f(x)g(x)> c(x-1)^2$ for all $x>0$. Since $f(x)g(x)$ is increasing while $c(x-1)^2$
is decreasing for $0<x\le1$, it is clear that $f(x)g(x)> c(x-1)^2$ for $0<x\le1$.
Since $f(x)$ and $g(x)$ are (operator) concave, it follows that
$f(x)g(x)/x^2=(f(x)/x)(g(x)/x)$ is decreasing for $x>0$. But $c(x-1)^2/x^2$ is
increasing for $x\ge1$, so that we have $f(x)g(x) > c(x-1)^2$ for $x\ge1$ as well.
The inequality shown above implies that
$$
M_g(\lambda_i,\lambda_j)>c{(\lambda_i-\lambda_j)^2
\over M_f(\lambda_i,\lambda_j)}
$$
for all $1\le i,j\le m$. Hence $\bal \pont , \pont \jobb - 
\< \pont ,\pont\>$ is strictly positive on $\iK$, and the latter
statement  follows from Lemma \ref{L:2}. \qed

Recall that (\ref{dyn}) is obtained by the choice $g(x)=(1+x)/2$ and $c=f(0)/2$.
Assume we put $c=f(0)/2$. Then (\ref{E:fo}) holds for a standard $f$ if
$$
g({x})\ge \frac{f(0)({x}-1)^2}{2f({x})}.
$$
In particular, $g(0)\ge 1/2$. The only standard $g$ satisfying this inequality
is $g(t)=(t+1)/2$. This corresponds to the {case where} the left-hand-side 
is the usual covariance.

Motivated by \cite{LZ1, WYD}, Kosaki \cite{Kosaki} studied the case when 
$f({x})$ equals to
$$
h_\beta({x})=\frac{\beta(1-\beta)({x}-1)^{2}}
{({x}^{\beta}-1)({x}^{1-\beta}-1)}.
$$
In this case $g({x})=h_\beta({x})$ is possible for every
$0 < \beta < 1$ if the constant $c$ is chosen properly. More generally, inequality
(\ref{E:fo}) holds for any standard $f$ and $g$ when the  constant $c$ is
appropriate. It follows from the lemma {below} that $c=f(0)g(0)$ is
good, see (\ref{E:gibi}).

\begin{lemma}
For every standard function $f${,}
$$
f(x) \ge f(0)\, |x-1|\,.
$$
\end{lemma}

\proof
The inequality is not trivial only if $f(0)>0$ and $x>1$, so assume these conditions.
Let $q(x_0)$ be the constant such that the tangent line to the graph of $f$ at 
the point $x_0>1$ has the equation
$$
y=f'(x_0) x + q(x_0).
$$
Since $f$ is (operator) concave one has $q(x_0) \ge f(0)$.
Using again (operator) concavity and symmetry one has
$$
f'(x_0) \geq \lim_{x \to + \infty} f'(x)= \lim_{x \to + \infty} 
\frac{f(x)}{x}=\lim_{x \to + \infty}f(x^{-1})=f(0)>0.
$$
This implies
$$
f(x_0)=f'(x_0)\cdot x_0 + q(x_0) \geq f(0)\cdot x_0+f(0) 
\geq f(0)\cdot x_0-f(0)= f(0)\cdot( x_0-1)
$$
and the proof is complete. \qed

The lemma gives the inequality
\begin{equation}\label{E:gibi}
f(x)g(x)\ge f(0)g(0) (x-1)^2
\end{equation}
for standard functions. If $f(0)>0$ and $g(0)>0$, then Theorem \ref{T:3}
applies.

Similarly to the proof of Theorem \ref{T:3}, one can prove that 
the right-hand-side of (\ref{E:fo}) is a monotone function of the variable $f$.
 
\begin{Thm}
Assume that $f,g:\bbbr^+\to \bbbr$ are standard functions. If
\begin{equation}\label{E:felt2}
\frac{c}{f(t)}\ge \frac{d}{g(t)}
\end{equation}
for some positive constants $c,d$ and $A_1,A_2, \dots ,A_m$ are self-adjoint 
matrices, then
\begin{equation}
\Det\biggl( \left[ c \,\gamma_D^f\left( \,[D,A_{i}], [D,A_{j}]\, \right)
\right]_{i,j=1}^m  \biggr)
\le
\Det\biggl( \left[d \,\gamma_D^g\left( \,[D,A_{i}], [D,A_{j}]\, \right)
\right]_{i,j=1}^m  \biggr)
\end{equation}
holds.
\end{Thm}

\end{document}